\documentclass[sigconf]{acmart}
\renewcommand\footnotetextcopyrightpermission[1]{} 

\AtBeginDocument{%
  \providecommand\BibTeX{{%
    \normalfont B\kern-0.5em{\scshape i\kern-0.25em b}\kern-0.8em\TeX}}}
\setcopyright{none}
\copyrightyear{}
\acmYear{}
\acmDOI{}
\acmConference[Document Intelligence Workshop at KDD]{The Second Document Intelligence Workshop at KDD}{2021}{Virtual Event}
\acmPrice{}
\acmISBN{}

\begin{document}


\title {Towards Semantic Search for Community Question Answering for Mortgage Officers}


\author{Amir Reza Rahmani}
\email{{amirrezar, linweil,  brianv,  colinbe,  srawat}@zillowgroup.com}
\author{Linwei Li}
\author{Brian Vanover}
\author{Colin Bertrand}
\author{Shourabh Rawat}
\affiliation{%
  \institution{Zillow Group Automation and AI}
  \streetaddress{1301 2nd Ave}
  \city{Seattle}
  \state{Washington}
  \country{USA}
  \postcode{98101}
}

\renewcommand{\shortauthors}{Rahmani and Li, et al.}

\begin{abstract}
Community Question Answering (CQA) has gained increasing popularity in many domains.  
Mortgage is a complex and dynamic industry, and a flexible and efficient CQA platform can potentially enhance the quality of service for mortgage officers significantly. We have built a dynamic CQA platform with a state of the art semantic search engine based on recent Natural Language Processing (NLP) techniques to dynamically and collectively capture and transfer the maturity and tribal knowledge of the more experienced workforce to less experienced ones. The search engine allows for both keyword and natural language queries and is based on a fine-tuned domain-adapted Sentence-BERT encoder linearly composed with a TF-IDF vectorizer, and reciprocal-rank fused with a BM25 vectorizer. Domain adaptation and fine-tuning is based on publicly available mortgage corpora. Evaluation is performed on an internally annotated dataset using standard information retrieval metrics such as normalized discounted cumulative gain (nDCG), precision/recall at n, mean reciprocal rank, and mean average precision (MAP). The results indicate that our hybrid, fine-tuned, domain-adapted search engine is a more effective approach in responding to the information needs of our mortgage officers compared to traditional search techniques. We aim to publish the internally-annotated evaluation and training datasets in the near future.

\end{abstract}



\maketitle

\section{Introduction}
A CQA forum combined with an advanced search engine that allows the users to also submit their queries in natural language could significantly boost the efficiency of operations in a fast-paced organization. In the mortgage industry, where operations are based on dynamic interactions with a large population of customers, a question answering platform will help ensure the consistency of the services offered, on-board and train new members, capture the tribal knowledge of the team, and also offer more versatile products and services to meet the buyers' needs.

With these motivations in mind, we have developed a CQA system for all the professionals who are involved in Zillow Group's mortgage operations. Our system consists of two major parts: a) a question-answering platform which allows the users to post questions, answer their peers' questions, up-vote and approve answers, and track their peers activities, b) a state of the art search engine that can search public mortgage resources as well as internal contents that our mortgage team produces using the QA platform.  

To build an effective system for our mortgage staff, the search engine needs to a) accept natural language as well as keyword queries,
 b) return highly relevant search results right at the top, and c) do so within a reasonable time. Hybrid search engines, i.e. search engines that perform semantic and keyword search, are a promising choice to satisfy these requirements. Many efforts have been made in this direction. For example, \cite{salesforce_covid, mass-etal-2020} implemented retrieval systems by combining Sentence-BERT\cite{sbert} and traditional retrieval models such as BM25\cite{bm25}. 

Our search engine uses a hybrid approach
to search over all the internal and external data. Our database consists of FAQ-answer pairs that are gathered by using publicly available FAQ pages of business related entities (external) and also questions and answers that our mortgage staff input into the forum (internal). When a user inputs a query in the search box, the query is matched against all the external FAQ-answer pairs as well as all the contents from our internal forum. To train and tune the search engine, 
we fine-tune a Sentence-BERT model and train a TF-IDF\cite{tfidf1986} and a BM25 model on all the FAQ-answer pairs. At run time, for an incoming query, the Sentence-BERT score for the question part of every FAQ-answer pair is linearly combined with the TF-IDF score on both question and answer parts of all FAQ-answer pairs. The resultant scores generate a ranking which is reciprocally fused with the BM25 ranking. 
The linear combination between the TF-IDF and Sentence-BERT score is controlled using a dampening factor which gives more weight on SBERT for longer queries. 
Although our search algorithm is inspired by \cite{salesforce_covid}, they differ in the following aspects: a) we use FAQ-answer pairs as our documents and we combine the contributions from both question and answer parts during indexing and inference, b) we use a dampening factor during inference based on the length of the query to control the degree of Sentence-BERT and TF-IDF combination, and c) our search engine is specifically fine-tuned to mortgage domain and is evaluated on a domain-specific dataset. 

Finally, the emphasis of this work is to develop a functioning system that translates algorithmic ideas into a novel industry context and to solve a practical business problem. The system is currently being utilized by our mortgage staff and with the user interface capturing feedback, we are collecting data to further improve the system and pragmatically develop the product road map.

\begin{figure*}[h]
  \centering
  \includegraphics[width=0.8\linewidth]{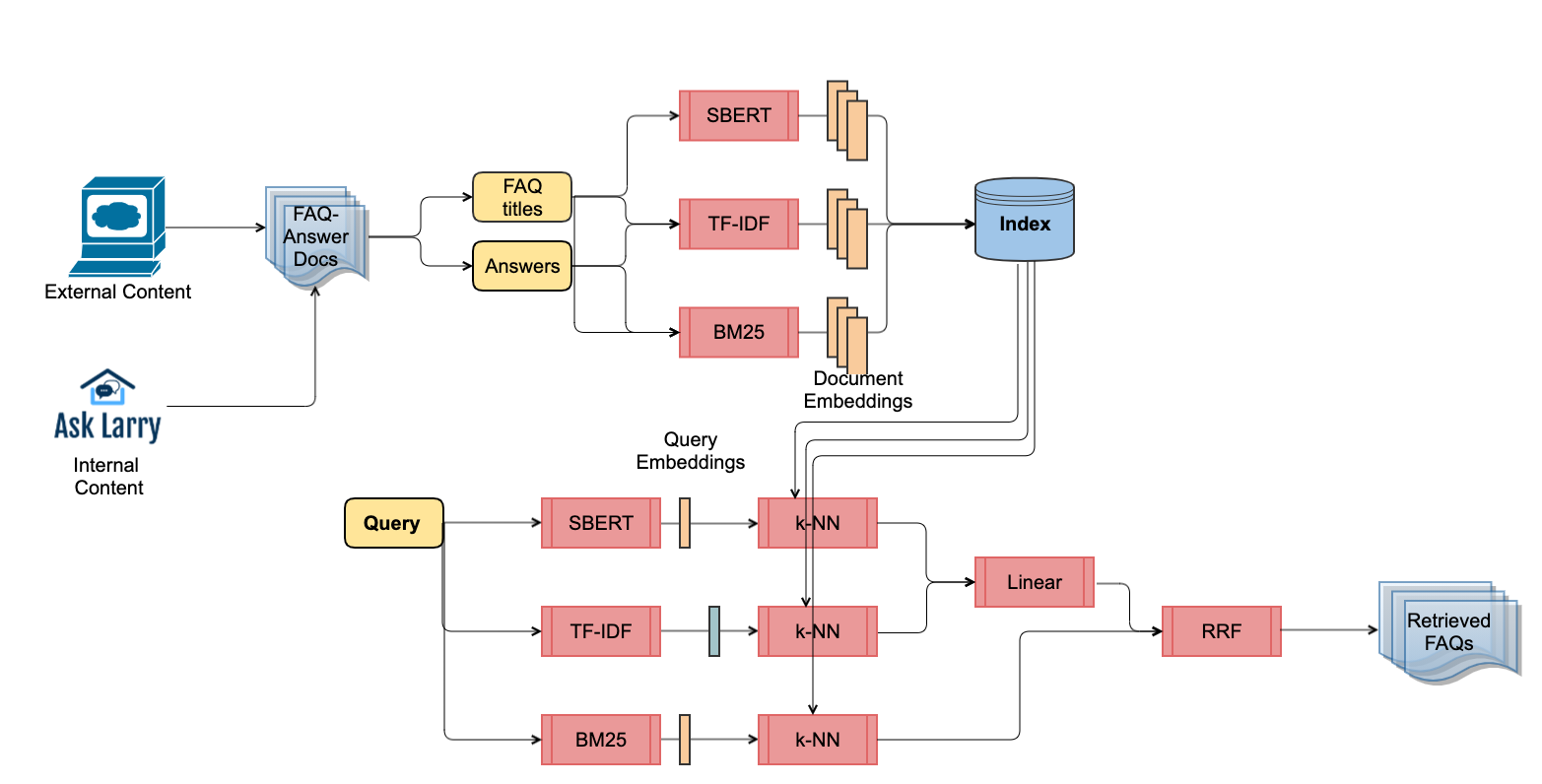}
  \caption{System Architecture}
  \Description{System Architecture}
  \label{fig:system_arch}
\end{figure*}

\section{Model}
The search engine (Figure \ref{fig:system_arch}) consists of two components, representation learning and retrieval. During representation learning, a preprocessed tokenized FAQ-answer pair $(q,a)$ is taken as the input and three embeddings are generated through a Sentence-BERT model and a TF-IDF vectorizer. The Sentence-BERT embedding $u$ only attends to the question part of the FAQ-answer pair and the TF-IDF vectorizer outputs $e^{q}$ for the question and $e^{a}$ for the answer part. During retrieval, a query $Q$ is provided and multiple FAQ-answer pairs $(q_1, a_1), \dots, (q_n, a_n) $ are returned as the retrieval result. The ranking of the retrieved FAQ-answer pairs is determined by a reciprocal rank fusion over the ranking produced by the aforementioned embeddings' cosine similarity scores and a BM25 ranking.
\subsection{Representation Learning}
\label{sec: Representation Learning}
\subsubsection{Sentence-BERT (SBERT)}
We have observed in our dataset that:
\begin{enumerate}
\item The question part of an FAQ-answer pair is often a good summary of the answer's topic.
\item FAQs are often grammatically complete sentences.
\item Similar concepts can be expressed in various lexical forms, e.g. loan and mortgage. 
\end{enumerate}

Transformer based language models are effective in capturing the semantics, however, they are also known to be slow during training and inference. Based on observations (1) and (2), SBERT model becomes a good fit for our task because it directly generates embeddings that cluster semantically similar sentences together without further processing needed during inference. The SBERT model uses a siamese network to capture the semantic textual similarity between sentence pairs and uses a triplet or a classification loss to guide the network. Their proposed way of using cosine similarity along with the siamese network significantly reduces the search time as each sentence is encoded only once whereas in the original BERT model, every pair needs to be fed in the network (separated by the <SEP> token). Furthermore, \cite{sbert} showed that BERT embeddings (either based on the <CLS> token or obtained by averaging the outputs) are not effective at capturing sentence semantics and are outperformed by GloVe embeddings \cite{glove}.

The training process involves two steps:
\paragraph{BERT finetuning} 
We perform language model fine-tuning on the bert-base model\cite{huggingface} by applying random masks over the sentences selected from the mortgage FAQ-answer pairs. 
\paragraph{SBERT training}
Based on the finetuned BERT model, we further train the SBERT model using two different approaches. The first approach is to train the model in a classification task as mentioned in \cite{sbert}. The SBERT model is trained on the SNLI and multi-genre NLI data \cite{snli, multigenre_nli}. 

The second approach is to use a triplet objective function, also discussed in \cite{sbert}. A triplet comprising an anchor sentence $s_a$, a positive sentence $s_p$, and a negative sentence $s_n$ are used as the inputs. The model generates embeddings $u_a, u_n, u_p$ for each of the three sentences and the objective is to maximize $\lVert u_a-u_n\lVert$ and minimize $\lVert u_a-u_p \lVert$. When constructing the triplets, the positive sentence is the sentence immediately following the anchor sentence; and the negative sentence is sampled from a random FAQ-answer pair. During experiments, we found that the model easily differentiates the negative sentences from the anchor sentences, failing to learn meaningful representations. We therefore construct another set of triplets by selecting negative sentences from the FAQ-answer pairs of the same category as the anchor sentence.

\subsubsection{TF-IDF}
We follow the standard TF-IDF computation to generate the embeddings of both question and answer parts in FAQ-answer pairs.

\subsection{Retrieval and Re-ranking}
During inference, the system retrieves an ordered list of documents that best matches the query $Q$. The initial ranking is produced by computing the similarity scores between the query embedding and the embeddings for question and answer parts of the FAQ-answer pair. The score is a weighted sum of cosine similarities computed over SBERT embeddings and TF-IDF embeddings:
\begin{align*}
F_\text{initial}(Q, (q, a)) &= [\alpha + (1-\alpha)*\zeta] * F_{\text{sbert}}(Q, (q, a)) \\
                            & + [(1-\alpha)*(1-\zeta)]* F_{\text{tfidf}}(Q, (q, a)) \\
F_{\text{sbert}}(Q, (q, a)) &= \text{cos}(u_Q, u_{FAQ}) \\
F_{\text{tfidf}}(Q, (q, a)) &= w * \text{cos}(e_Q, e^{q}_{FAQ}) + (1-w) * \text{cos}(e_Q, e^{a}_{FAQ}) \\
\zeta &= exp(\frac{1-len(Q)}{\beta})
\end{align*}
where $F_{\text{sbert}}$ and $F_{\text{tfidf}}$ are the similarity scores based on SBERT and TF-IDF, $u_Q$, $u_{FAQ}$ are SBERT embeddings for the query and the question part of the FAQ-answer pair, respectively. $e_{Q}$ is the TF-IDF embedding for the query, $e^{q}_{FAQ}, e^{a}_{FAQ}$ are TF-IDF embeddings for the question and answer parts of the FAQ-answer pair, respectively. $\zeta$ is a damping factor which favors TF-IDF similarity for shorter queries because they are more likely to be keywords, and $\alpha$, $w$ and $\beta$ are hyperparameters. $\alpha$ controls the combination of SBERT and TF-IDF similarity scores, $w$ controls the combination of TF-IDF similarity scores between the query-question and query-answer parts, and $\beta$ controls the damping factor.

The second rankings is produced by a BM25 ranker, with the same weight $w$ combining the question and the answer parts as in TF-IDF. In the end, both rankings are combined using reciprocal rank fusion as follows:
$$\text{RRF}(Q, (q, a)) = \frac{1}{k+R_\text{SBERT+TFIDF}}+\frac{1}{k+R_\text{BM25}}$$
where $R_\text{SBERT+TFIDF}$ is the initial ranking of the FAQ-answer pairs produced by SBERT and TF-IDF, $R_\text{BM25}$ is the BM25 ranking, and $k$ is a hyperparameter. 

\section{Experimental Setup}
\subsection{Datasets}

\subsubsection{Training Dataset}
The training dataset consists of around 6000 FAQ-answer pairs acquired from publicly available mortgage data. This dataset is used for BERT domain language model fine-tuning. The question part of the FAQ is usually less than two sentences whereas the answer part can range from a single paragraph to a few pages. About \text{80\%} of the FAQs have categories associated to them (e.g. appraisal, mortgage forbearance, etc). These categories are used when creating triplets for training SBERT (see \ref{sec: Representation Learning}).

\subsubsection{Evaluation Dataset}
We have picked the top 16 queries that our loan officers most frequently search. For each query, we collect the top 50 FAQ-answer pairs based on TF-IDF, BM25, and SBERT, respectively, keeping only the unique pairs. Therefore, for each query, we have at most 150 FAQ-answer pairs to label. We use three labels: 0 (irrelevant), 1 (somewhat relevant), 2 (relevant). See Appendix \ref{appendix:dataset} for more details on the dataset.

\subsection{Evaluation} 
\label{section_eval}
We evaluate our search engine against the labelled data from the previous step. We use TREC-EVAL \footnote{\url{https://trec.nist.gov/trec_eval/}} to evaluate our search engine based on the following metrics:

\begin{itemize}
\item Mean Average Precision (MAP, MAP@5, MAP@10)
\item Mean reciprocal rank 
\item Normalized Discounted Cumulative Gain (nDCG@5, \\
nDCG@10)
\item P@5, P@10, 
\item Recall@5, Recall@10 
\end{itemize}

The metrics are selected based on the original system requirements, i.e. returning the most relevant results right at the top; we need the top five retrieved results to contain as much relevant information as possible. 

We will compare the following approaches:
\begin{itemize}
    \item SBERT
    \item TF-IDF
    \item BM25
    \item TF-IDF combined with SBERT
    \item Reciprocal Rank Fusion (RRF)
\end{itemize}
The grid search is performed over $w$ and $\alpha$, which represent the linear factor controlling the combination of FAQ and answer contributions in TF-IDF and BM25, and the combination of TF-IDF and SBERT contributions, respectively. They both range between 0 and 1 with a step size of $0.1$. We set $\beta$ to $3$ for now but we intend to include it in the grid search in the future.


\begin{table*}[h]
\small
  \caption{Comparison of different algorithms on internal evaluation dataset}
  \label{tab:eval_results}
  \begin{tabular}{lcccccccc}
    \toprule
    Algorithm & Mean Reciprocal Rank & Recall@5 & Recall@10 & nDCG@5 & nDCG@10 & MAP & MAP@5 & MAP@10  \\
    \midrule
    TF-IDF & 0.8507 & 0.2370 & 0.3364 & 0.6160 & 0.5517 & 0.4485 & 0.2289 & 0.3014\\
    BM25& 0.8221 & 0.2431& \textbf{0.3671} & 0.6579 & \textbf{0.5895} & \textbf{0.4923}& 0.2288 & \textbf{0.3283} \\
    Ours: SBERT + TF-IDF& \textbf{0.9375}&  0.2583 & 0.3597 & \textbf{0.6625}& 0.5755 & 0.4439 & \textbf{0.2518}  & 0.3182\\
    Ours: SBERT + TF-IDF + BM25& 0.8828 & \textbf{0.2589} & 0.3586 & 0.6600 & 0.5816 & 0.4874 & 0.2308 & 0.3197 \\
  \bottomrule
\end{tabular}
\end{table*}

\begin{table*}[h]
\small
  \caption{Non-trivial queries matched by the semantic search}
  \label{tab:hard_queries}
  \begin{tabular}{ll}
    \toprule
    Query&Relevant FAQ\\
    \midrule
    Can we originate a loan for a home on the market? & Can a property be refinanced if it is currently listed for sale?\\
    Minimum size for manufactured house?& What are the requirements for a living unit?\\
    What credit counseling advice can we give borrowers?& What resources can I provide to applicants to help improve their credit score?\\
  \bottomrule
\end{tabular}
\end{table*}


\begin{table*}[ht!]
\small
  \caption{Comparison of SBERT models performance on the evaluation data}
  \label{tab:sbert_comp}
  \begin{tabular}{lcccccccc}
    \toprule
    Model & Mean Reciprocal Rank & P@5 & P@10 & Recall@5 & Recall@10 & nDCG & MAP\\
    \midrule
    SBERT-base (Base BERT + NLI) & 0.6073 & 0.3500 & 0.2250 & 0.1328 & 0.1607 & 0.2995& 0.1580\\
    SBERT-mor-nli (fine-tuned BERT + NLI)
    &\textbf{0.6646}&\textbf{0.4250}  &  
    \textbf{0.2437} & \textbf{0.1559} & \textbf{0.1790} & \textbf{0.3590} & \textbf{0.1743}\\
    SBERT-mor-triplet (fine-tuned BERT + mortgage triplets) & 0.5107 & 0.2250 & 0.1562 & 0.0572 & 0.0780 & 0.1861& 0.0738\\
  \bottomrule
\end{tabular}
\end{table*}

\section{System Architecture}
Our system architecture (Figure \ref{fig:system_arch}) is inspired by \cite{salesforce_covid}. The collected FAQ-answer pairs are stored in a database table. Each document contains an FAQ part (or title) and an answer part. We index them using SBERT, TF-IDF, and BM25 and store the indexes in a database. The indexes are stored as matrices containing document (title/answer) embeddings. When a new query is inserted in the forum, it is also indexed by SBERT, TF-IDF, and BM25. All the document indexes are loaded and the query index is compared against them. Using cosine similarity based KNN, we find the top $N$ from each of the indexes. We then linearly combine the SBERT and TF-IDF results according to the equations described above and combine the resultant ranking with BM25 ranking using reciprocal rank fusion (RRF). The top $M$ results based on the RRF are surfaced to the user. In our system, we set $N = 200$ and $M = 50$.

\section{Results}

\subsection{Search Engine Evaluation Results}
We perform grid search on the hyperparameters $\alpha$ and $w$ and also on the retrieval and ranking approaches mentioned in \ref{section_eval}. 

Table \ref{tab:eval_results} displays our metrics for different retrieval and ranking approaches. For each approach, the reported metric corresponds to the model with tuned hyperparameters. The SBERT model reported in the table is trained using the classification loss based on the finetuned BERT (SBERT-mor-nli in Table \ref{tab:sbert_comp}). The grid search results indicate that inclusion of SBERT either in TF-IDF\_SBERT or RRF creates a more efficient ranking for the top five retrieved results, and therefore, satisfies our requirement. This is specifically very visible in mean reciprocal rank, nDCG@5 and MAP@5. BM25 or a combination of BM25 and TF-IDF outperform the other models when ranking is extended beyond the top five retrieved results. For example, BM25 outperforms when comparing Recall@10, nDCG@10, MAP, and MAP@10.  

There are some queries in our evaluation set that show clearly that inclusion of SBERT helps in capturing the relevant documents when there is minimal lexical overlap where the keyword based methods (BM25 and TF-IDF) fail to capture (see Table \ref{tab:hard_queries}).

\subsection{Sentence-BERT Models Comparisons}
We compare three SBERT models. The first one is the base model, which is used as benchmark. This model uses regular BERT model and is trained on SNLI and multi-genre NLI data (\cite{snli, multigenre_nli}) and is made available by the SBERT developers. The second model is built based on the mortgage fine-tuned BERT and is trained on SNLI and NLI data. The third model is based on the mortgage fine-tuned BERT and is trained on triplet sentences generated based on the approach explained in Section \ref{sec: Representation Learning}. We compare the performance of the three models (Table \ref{tab:sbert_comp}) on our evaluation data using the metrics discussed in Section \ref{section_eval}.

Table \ref{tab:sbert_comp} suggests that the SBERT model based on mortgage fine-tuned BERT and trained on NLI data outperforms both the benchmark model and the SBERT model trained on triplets. Therefore, we use this model in the search engine. This shows that fine-tuning BERT on the mortgage domain is effective for the downstream task. The reason that the third model did not yield better results is that finding difficult triplets to challenge the model into learning small nuances between close answers is itself a challenging task and requires further effort and a more effective strategy. Furthermore, the training dataset is small and more training data is required for the third model to outperform the second one.

\section{Discussion}
We have implemented a question answering system with a state of the art hybrid search engine for the mortgage domain. The system is customized to support and assist the mortgage staff at Zillow Group. We are currently measuring the impact on the business operations. Future directions mostly include but are not limited to:

\begin{itemize}
    \item Collect more query-candidate pairs annotations.
    \item Use larger mortgage corpora for fine-tuning the models.
    \item Our CQA system allows for collection of other kinds of data such as votes on questions and answers, whether an answer is approved by the original poster, the association of user-answer and user-question, or tags assigned to each question. The integration of such signals remains an open research question \cite{radlinski2008does, xue2004optimizing}.
    \item Adopt supervised training approaches such as Learning to Rank \cite{learning_to_rank} based on the collected user data. 
\end{itemize}


\bibliographystyle{ACM-Reference-Format}
\bibliography{sample-base}

\appendix

\section{Evaluation Dataset Statistics}
\label{appendix:dataset}
The evaluation dataset contains 16 queries and 1740 FAQ-answer pairs. Table \ref{tab:statistics} lists the summarized statistics for our evaluation dataset. Table \ref{tab:common_queries} shows the 16 queries that are used for evaluation. We also present an example of an FAQ-answer pair below:

\textbf{FAQ:} "How can I determine if a manufactured home is eligible for FHA financing?"

\textbf{Answer:} "Manufactured homes may be legally classified as real property or personal property based on how they are titled in accordance with state law. In general, manufactured homes can only be classified as real property if they are permanently affixed to the land..." (truncated)

\begin{table}[h]
\small
  \caption{Statistics of the evaluation dataset}
  \label{tab:statistics}
  \begin{tabular}{lr}
    \toprule
    Avg. (word) length of questions & 9.38 \\
    Avg. (word) length of answers & 174.69 \\
    Avg. number of FAQ-answer pairs per query & 108.75 \\
    Avg. number of relevant FAQ-answer pairs per query & 7.00 \\
    Avg. number of partially relevant FAQ-answer pairs per query & 8.44 \\
    Avg. number of non-relevant FAQ-answer pairs per query & 93.31 \\
    \bottomrule
\end{tabular}
\end{table}

\begin{table}[H]
\small
  \caption{Most commonly searched mortgage queries}
  \label{tab:common_queries}
  \begin{tabular}{p{\linewidth}}
    \toprule
Is there a minimum square footage requirement for a home to be eligible for FHA financing? \\
What's the maximum DTI ratio for a conventional loan? \\
Is manual underwriting allowed? \\
Can we originate a loan for a home on the market?\\
Do I need to collect reserves for a second home?\\
Can I give loan to a non US Citizen?\\
Can I give loan to a customer who has late payments?\\
Customer has a judgement, can I do the loan?\\
Can I do past-due, collection, and charge-off of non-mortgage accounts?\\
Customer pays alimony/child support, does that count against the DTI?\\
Customer only has 9 payments left on their car. Can I exclude it?\\
Customer has student loan, but the credit report says zero for payment?\\
What do I do with open accounts? Amex\\
He's seasonally employed, can I use that income?\\
She is starting a new job, what documents do I need?\\
Customer has foreign income. Can we use that?\\
  \bottomrule
\end{tabular}
\end{table}

\end{document}